\documentclass[conference]{IEEEtran}  									  

\usepackage{epsfig,cite}
\usepackage{subfig}
\usepackage{graphicx}
\usepackage{color}
\usepackage{amssymb,amsmath,mathtools, mathrsfs,amsthm}
\usepackage[mathscr]{euscript}
\usepackage{graphicx}
\usepackage{epstopdf}
\usepackage{caption}
\usepackage{tikz}
\graphicspath{{images/}}
\usetikzlibrary{arrows}

\theoremstyle{remark} \newtheorem{remark}{Remark}

\theoremstyle{remark}

\newcommand{\realSet}{\mathcal{R}}

\renewcommand{\vec}[1]{\mathbf{#1}}

\newcommand{\changeColor}{black}
\newcommand{\editColor}{black}

\IEEEoverridecommandlockouts
\title{Ranking Causal Influence of Financial\\ Markets via Directed Information Graphs
\thanks{
\noindent The work was partly supported by the NSF Center for Science of Information (CSoI) under grant NSF-CCF-0939370.
}
}

\author{Theo Diamandis, Yonathan Murin, and Andrea Goldsmith \\
{\small Department of Electrical Engineering, Stanford University, Stanford CA, USA}
\vspace{-0.35cm}
}

\begin{document}

\maketitle
\thispagestyle{empty}
\pagestyle{empty}

\begin{abstract}

A non-parametric method for ranking stock indices according to their mutual causal influences is presented. Under the assumption that indices reflect the underlying economy of a country, such a ranking indicates which countries exert the most economic influence in an examined subset of the global economy.
The proposed method represents the indices as nodes in a directed graph, where the edges' weights are estimates of the pair-wise causal influences, quantified using the directed information functional. 
This method facilitates using a relatively small number of samples from each index.
The indices are then ranked according to their net-flow in the estimated graph (sum of the incoming weights subtracted from the sum of outgoing weights).
Daily and minute-by-minute data from nine indices (three from Asia, three from Europe and three from the US) were analyzed.
The analysis of daily data indicates that the US indices are the most influential, which is consistent with intuition that the indices representing larger economies usually exert more influence. Yet, it is also shown that an index representing a small economy can strongly influence an index representing a large economy \textcolor{\editColor}{if the smaller economy is indicative of a larger phenomenon.}
Finally, it is shown that while inter-region interactions can be captured using daily data, intra-region interactions require more frequent samples.


\end{abstract}

\vspace{-0.15cm}
\section{Introduction} 

\vspace{-0.05cm}
The causal structure that underlies the global economy governs propagation of risk, economic shocks, and economic strength. As a result, this structure is important to investors and regulators alike, and much work in this field examines financial time-series to try to infer the propagation of information around the global economy \cite{Billio2012,Sandoval2015,Dimpfl2014}. Some of the existing methods for quantifying causal interactions provide accurate measures of this casual structure when the data conforms closely to an estimated model. 
For instance, the works \cite{Billio2012, Ratanapakorn2002} applied analysis based on Granger Causality (GC) \cite{Granger69},  \textcolor{\changeColor}{which assumes a specific model} for the interactions between the observed time-series.
On the other hand, when there is a mismatch between the observed time-series and the assumed model, model-based methods may provide poor inference results. 
This \textcolor{\changeColor}{consideration} motivates designing data-driven approaches for inferring causal influences between financial time-series, building upon the fact that non-parametric methods can pick up interactions that model-based estimators fail to uncover. 
In this work, we consider a set of stock indices (from a subset of the world's stock exchanges), and rank these indices \textcolor{\changeColor}{in terms of the} causal influence \textcolor{\changeColor}{they exert on} the rest of the members of the analyzed set. It is reasonable to expect that the indices representing countries with larger economies will exert more influence than indices representing countries with
smaller economies. Yet, we show that this is not always the case, and in scenarios \textcolor{\editColor}{where a small economy reflects the behavior of a larger phenomenon,} an index representing a country with a relatively small economy can strongly influence an index representing a country with a very large economy.
Under the assumption that indices reflect the underlying economy of a country, this ranking indicates which countries exert the most economic influence in the examined subset of the global economy.

Non-parametric inference of the causal influence structure underlying a set of observed time-series was studied in \cite{Rahimzamani16} where it was shown that, under mild assumptions, this structure can be accurately reconstructed from the observed time-series if the number of available samples in each time-series is large enough. Similarly to the ideas presented in \cite{Quinn2011}, \cite{Rahimzamani16} also proposed to evaluate the causal influence between two time-series {\em when statistically conditioning on the rest of the observed time-series}. As a measure for causal influence, \cite{Rahimzamani16} used the directed information (DI) functional (which is closely related to the transfer entropy (TE) functional). 
\textcolor{\changeColor}{Noting that the DI must be estimated from the observed time-series, conditioning on the rest of the time-series significantly enlarged the effective state-space in the estimation problem}; even when the number of time-series is moderate, this task requires a huge number of samples.
As the number of available samples in the problem studied in the current work is relatively small, we do not try to infer the underlying causal structure as in \cite{Rahimzamani16}. Instead of estimating the DI \textcolor{\changeColor}{by} conditioning on the rest of the time-series, we estimate the non-conditioned pair-wise DI, which requires significantly fewer samples. We then use the estimated values as \textcolor{\changeColor}{edge} weights in a graph, where the nodes correspond to the time-series, and the ranking is performed by processing the estimated graph.

Several studies applied TE (or equivalently DI) \textcolor{\changeColor}{for} non-parametric inference of causal interactions underlying financial time-series (see \cite{Dimpfl2014,Sandoval2015,Kwon2008,Marschinski2002,Mantegna1999} and references therein).  Yet, the approach we propose in the current work is different in the following aspects: First, previous works quantized the observed time-series as part of estimating the TE. We, on the other hand, use an estimator based on the $k$-nearest-neighbor ($k$-NN) principle, allowing for a continuous input alphabet. Second, previous works placed strict assumptions on the length of the memory of the observed time-series, while we estimate this length from the data. 
Third, \textcolor{\changeColor}{some} previous works neglected differing trading hours of daily data (over indices from different regions, e.g., Asia and the US). As this approach ignores the time difference between different markets, it may \textcolor{\editColor}{show interactions that are, in fact, non-causal due to time differences}. To account for this challenge, we \textcolor{\editColor}{interlace} the analyzed time-series to reflect the differences in trading hours and days. \textcolor{\editColor}{This interlacing method is discussed in Section \ref{subsec:fin_preProc}.}

In this work we analyze nine indices from three regions of the world sampled at a daily interval. \textcolor{\changeColor}{We also analyze} the subset of the three indices from Europe sampled at a one minute interval. To quantify the pair-wise causal influence between a pair of indices, we first remove time points for which one of the indices was not calculated, then we offset sample points to reflect the correct time of trading in different counties, and finally we estimate the DIs, obtaining a directed graph. Motivated by the techniques of \cite{TrackingEpileptic, Malladi2016}, which studied the problem of localizing the seizure-onset-zone in epileptic patients, we rank the time-series according to the net-flow (sum of the incoming weights subtracted from the sum of outgoing weights) of their corresponding nodes in the graph. Our results indicate which indices exert the most influence, which indices are the most influenced, and how this influence propagates through the examined subset of the global market. 

The rest of this paper is organized as follows: Section \ref{sec:probFormul} formally defines the problem. Section \ref{sec:proposedMethod} introduces the proposed inference approach. Section \ref{sec:methods} describes the analyzed data and the applied pre-processing. Section \ref{sec:Results} presents the results of our analysis, and Section \ref{sec:conclusion} concludes the paper.

{\bf {\slshape Notation}:} We denote random variables (RVs) by upper case letters, $X$, and their realizations with the corresponding lower case letters. We use the short-hand notation $X_1^n$ to denote the sequence $\{X_1,X_2,\dots,X_n\}$. We denote random processes using boldface letters, e.g., $\mathbf{X}$, while matrices are denoted by sans-serif font, e.g., $\mathsf{G}$. We denote sets by calligraphic letters, e.g., $\mathcal{S}$, where $\realSet$ denotes the set of real numbers. $f_X(x)$ denotes the probability density function (PDF) of a continuous RV $X$ on $\realSet$, and $\log(\cdot)$ denotes the natural basis logarithm. Finally, we use $h(\cdot)$ and $I(\cdot;\cdot)$ to denote differential entropy and mutual information as defined in \cite[Ch. 8]{ElementsOfInformationTheory}.


\section{Problem Formulation} \label{sec:probFormul}

Let $\{\vec{X}_l\}_{l=1}^L$ be a sequence of discrete-time continuous-amplitude random processes, and let $X_{l,1}^N \in \realSet^N$ be an $N$-length sample path of $\vec{X}_l$. We assume that the processes $\{\vec{X}_l\}_{l=1}^L$ are stationary, ergodic, and Markovian in the observed sequences. Stationarity implies that the statistics of a random process are constant throughout the observed sequence. Ergodicity ensures that the observed sequence truly represents the underlying process, and it converges over a long enough number of samples. Markovity implies that the observed signals have finite memory and is a common assumption in modeling real systems. We formulate the Markovity of the sequence $X_{j,1}^N, j \mspace{-3mu} \in \mspace{-3mu} \{1,2,\dots,L \}$, with order $M_j$, via:
\vspace{-0.1cm}
\begin{align}
	f_{X_j} \mspace{-5mu} \left( \mspace{-2mu} x_{j,n}| \{ X_{l,1}^{n-1} \mspace{-2mu} \}_{l=1}^L \mspace{-2mu} \right) \mspace{-3mu} = \mspace{-3mu} f_{X_j} \mspace{-5mu} \left( \mspace{-2mu} x_{j,n}| \{ X_{l,n-M_j}^{n-1} \mspace{-2mu} \}_{l=1}^L \mspace{-2mu} \right) \mspace{-3mu}, \label{eq:markovAssump}
\end{align}

\vspace{-0.1cm}
\noindent \noindent for all $n \mspace{-3mu} > \mspace{-3mu} M_j$. In the following we also assume that $M_j$ is the Markov order for any subset $\mathcal{L} \subset \{1,2,\dots,L\}$, namely:
\vspace{-0.1cm}
\begin{align}
	f_{X_j} \mspace{-5mu} \left( \mspace{-2mu} x_{j,n}| \{ X_{l,1}^{n-1} \mspace{-2mu} \}_{l \in \mathcal{L}} \mspace{-2mu} \right) \mspace{-3mu} = \mspace{-3mu} f_{X_j} \mspace{-5mu} \left( \mspace{-2mu} x_{j,n}| \{ X_{l,n-M_j}^{n-1} \mspace{-2mu} \}_{l \in \mathcal{L}} \mspace{-2mu} \right) \mspace{-3mu}, \label{eq:markovAssumpSubset}
\end{align}

\vspace{-0.1cm}
\noindent for all $n \mspace{-3mu} > \mspace{-3mu} M_j$. Next, we define the set $\mathcal{A}_j \in \{1,2,\dots,L\}$ to be the {\em minimal} set such that:
\vspace{-0.1cm}
\begin{align}
	f_{X_j}\left( x_{j,n}| \{ X_{l,1}^{n-1} \}_{l=1}^L \right) = f_{X_j}\left( x_{j,n}| \{ X_{l,n-M_j}^{n-1} \}_{l \in \mathcal{A}_j} \right). \label{eq:causalSet}
\end{align}

\vspace{-0.1cm}
\noindent We further assume that the set $\mathcal{A}_j$ is {\em unique} for all $j$. 
We say that there is causal relationship from $X_i^N$ to $X_j^N, j \neq i$, iff $i \in \mathcal{A}_j$. These causal relationships can also be encapsulated using a directed graph $\sf{G}$, where nodes correspond to processes, and there is a direct edge from node $i$ to node $j$ if $i \in \mathcal{A}_j$. The edges' weights in such a graph may quantify the level of causal influence between the respective processes.\footnote{Note that quantification of statistical causal influence among time-series is a long standing problem in signal processing \cite{Hlayackova07}.}

Our objective in the current work is to {\em rank} the processes $\{\vec{X}_l\}_{l=1}^L$ {\em in terms of their causal influence} within the network (graph) $\sf{G}$. In particular, we would like to understand which processes are the most influential in terms of causal influence on the rest of the processes. We note here that the structure of the graph $\sf{G}$ is {\em unknown}, and all we observe are the sequences $X_{l,1}^N$. While \cite[Thm. IV.1]{Rahimzamani16} implies that under mild assumptions the structure of the graph $\sf{G}$ can be accurately estimated, it requires $N$ to be very large, as discussed in more detail in \cite[Section V]{Rahimzamani16}. Therefore, in this work, \textcolor{\changeColor}{we focus} on designing an algorithm for ranking the processes $\{\vec{X}_l\}_{l=1}^L$ when a {\em relatively small} number of samples is available. 

\vspace{-0.1cm}
\section{The Proposed Approach} \label{sec:proposedMethod}

\begin{figure}
	\captionsetup{font=footnotesize}
	\begin{center}
		\includegraphics[width=0.96\columnwidth,keepaspectratio]{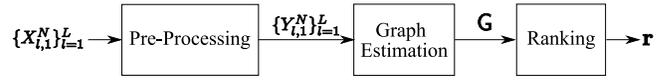}
	\end{center}
	\caption{\label{fig:BlockDiagram} {\bf Block diagram of the proposed method}. $\{ X_{l,1}^N\}_{l=1}^N$ are the observed sequences. $\{ Y_{l,1}^N\}_{l=1}^N$ are the processed sequences. $\mathsf{G}$ is the estimated graph, and $\mathbf{r}$ is a vector of rankings.}
	\vspace{-0.4cm}
\end{figure}

Fig. \ref{fig:BlockDiagram} provides a block diagram of the proposed approach. The input to the proposed method is the sequences $\{X_{j,1}^N\}$. First, the algorithm applies pre-processing on these sequences resulting in the sequences $\{Y_{j,1}^N\}$. 
The specific pre-processing applied to financial data is described in Subsection \ref{subsec:fin_preProc}. 
Then, the algorithm estimates the graph $\sf{G}$, where the weight of the edge from node $i$ to node $j$, i.e., $\mathsf{G}_{i,j}$, is the (estimated) pair-wise DI from $Y_{i,1}^N$ to $Y_{j,1}^N$, without conditioning on the rest of the sequences. Thus, this weight aims at quantifying the causal influence from the $i^{\text{th}}$ to the $j^{\text{th}}$ process. The method for estimating the DI is discussed in the next subsection. 
Finally, for each node in the graph, the sum of the incoming weights is subtracted from the sum of outgoing weights, and \textcolor{\changeColor}{we rank the nodes (processes)} according to this metric. 
In \cite{Malladi2016, TrackingEpileptic, MurinPCB} this approach was denoted by net-flow, and was shown to be effective for ranking nodes in a causal influence graph. In the case that all the weights are equal, net-flow ranks according to the degree centrality of the nodes. The work \cite{MurinPCB} also proposed a ranking method based on the PageRank algorithm \cite{PageRankBeyondTheWeb}. For the analysis of financial data, we observed that the method based on PageRank yields results similar to the net-flow approach.


\subsection{Estimating the Pair-Wise Directed Information} \label{subsec:estDI}

The DI from $Y_{i,1}^N$ to $Y_{j,1}^N$ is defined as \cite{Jiao2013}:
\vspace{-0.1cm}
\begin{align}
\label{eq:directedInformation}
I(Y_{i,1}^N \rightarrow Y_{j,1}^N) \triangleq \sum_{n=1}^N I(Y_{i,1}^{n-1};Y_{j,n}|Y_{j,1}^{n-1}).
\end{align}

\vspace{-0.1cm}
\noindent Note that in contrast to the definition used in \cite{Jiao2013}, we assume that only the {\em past} of the $i^{\text{th}}$ sequence influences its next sample. The expression \eqref{eq:directedInformation} depends on $N$, which motivates considering the DI {\em rate} between the two processes:
\vspace{-0.1cm}
\begin{align}
I(\vec{Y}_i \rightarrow \vec{Y}_j) \triangleq \lim_{N\rightarrow \infty} \frac{1}{N} I(Y_{i,1}^N \rightarrow Y_{j,1}^N).
\label{eq:DIRate}
\end{align}

\vspace{-0.1cm}
\noindent In the following, in addition to the Markovity assumption, we assume that $|h(Y_{j,1})| \mspace{-3mu} < \mspace{-3mu} \infty$, and that $|h(Y_{j,M_{j+1}}|Y_{j,M_j},Y_{i,M_j})| \mspace{-3mu} < \mspace{-3mu} \infty$. Under these assumptions, \cite{Malladi2016} showed that the DI rate exists and is equal to
\vspace{-0.1cm}
\begin{align}
	I(\vec{Y}_i \rightarrow \vec{Y}_j) = I(Y_{i,n-M_j}^{n-1};Y_{j,n} | Y_{j,n-M_j}^{n-1}), \quad n>M_j. \label{eq:DIRateSimp}
\end{align}

\vspace{-0.1cm}
\noindent Thus, the DI rate in \eqref{eq:DIRateSimp} can be viewed as quantifying the improvement in prediction of the next sample of $\vec{Y}_j$, given the past samples of $\vec{Y}_j$, compared to prediction when the past of $\vec{Y}_i$ is also available. 
To estimate the pair-wise DI, we use the $k$-NN estimator proposed in \cite{TransferEntropyInNeuroscience} and discussed in detail in \cite{knnEstDI}. 
We emphasize that using an estimator based on the $k$-NN principle allows us to use an effectively-continuous alphabet, whereas the works \cite{Jiao2013,Kwon2008,Sandoval2015,Marschinski2002,Dimpfl2013,Jizba2012} quantize the observed sequences into only a few bins to estimate empirical probability distributions.

It can be observed that \eqref{eq:DIRateSimp} is a function of $M_j$, which is in general unknown and must be estimated from the observed sequences. \cite{knnEstDI} \textcolor{\changeColor}{shows} that the performance of the $k$-NN estimator strongly depends on the estimated $M_j$. 
A possible approach for estimating $M_j$ is to choose the value that facilitates the best prediction of future samples (see \cite{Murin17BME} and references therein). Specifically, consider the set $\mathcal{M}_j$ of candidate Markov orders. $\hat{M}_j$ is estimated to be the value in this set that minimizes the average $\ell_2$ loss in predicting the next sample of $Y_{j,1}^N$ from $\hat{M}_j$ previous samples. 
While in \cite{TransferEntropyInNeuroscience} it was proposed to use the prediction method of \cite{Ragwitz2002}, i.e., to use $k$-NN prediction of the next sample in $Y_{j,1}^N$ based on the past samples of $Y_{j,1}^N$, this approach ignores the dependency between $Y_{j,1}^N$ and $Y_{i,1}^N$.
To account for this dependency, we predict the next sample of $Y_{j,1}^N$ based on the past $\hat{M}_j$ samples of both $Y_{j,1}^N$ and $Y_{i,1}^N$. Numerical simulations indicate that this approach leads to higher accuracy in estimating the DI \cite{knnEstDI}. 

\begin{remark}
	Previous studies, e.g., \cite{Marschinski2002,Billio2012,Kwon2008}, used an arbitrary value (usually $M_j=1$ for all estimated values) in quantifying the causal influence. While we also observed that for the analyzed financial time-series the values of $M_j$ are usually small, the proposed estimation procedure showed that they may vary between analyzed sequences, \textcolor{\editColor}{as shown in Table \ref{tbl:EstMarkovOrders}}. 
\end{remark}

\begin{remark}
	The $k$-NN prediction is known to have a relatively low computational complexity, and it can be combined with the nearest-neighbor calculations executed as part of estimating the DI. Moreover, in \cite{Murin17BME} it is shown that when $M_j$ is relatively small, $k$-NN has very good performance. On the other hand, when $M_j$ is large, $k$-NN suffers from the curse of dimensionality \cite{friedman2001elements}; \textcolor{\editColor}{when the dimensionality of the input grows, the fixed-size training data covers an exponentially decreasing fraction of the input space, resulting in poor performance for a high-dimensional input.} 
\end{remark}

\subsection{A Test Scenario}

\textcolor{\editColor}{Before discussing the financial data, we apply the proposed method to a simple test scenario to verify that this approach correctly ranks nonlinear simulated processes.}

Let $W_{l,n} \mspace{-3mu} \sim \mspace{-3mu} \mathcal{N}(0,1), l \mspace{-3mu} = \mspace{-3mu} 1,2,3,4$, be independent over time and independent of each other. 
For our test scenario, we consider 2000 samples from the non-linear causal network consisting of the following time-series:
\vspace{-0.1cm}
\begin{align}
	X_{1,n} \mspace{-3mu} &= \mspace{-3mu} W_{1,n}, && X_{2,n} \mspace{-3mu} = \mspace{-3mu} X_{1,n-1}^2 \mspace{-3mu} + \mspace{-3mu} X_{1,n-2}^2 \mspace{-3mu} + \mspace{-3mu} W_{2,n}, \nonumber \\
	X_{3,n} \mspace{-3mu} &= \mspace{-3mu} X_{2,n-1} \mspace{-3mu} + \mspace{-3mu} W_{3,n}, && X_{4,n} \mspace{-3mu} = \mspace{-3mu} X_{1,n-2} \mspace{-3mu} + \mspace{-3mu} W_{4,n}.
\end{align}

\vspace{-0.1cm}
\noindent \textcolor{\changeColor}{We rank} the nodes according to their incoming and outgoing flow in the estimated graph. Clearly, the most influential process is $\vec{X}_1$, and the second in the ranking vector should be $\vec{X}_2$. We repeated this experiment 24 independent times and in each of the trials the method described above resulted in the ranking $\vec{r} = [1,2,4,3]$, namely, $\vec{X}_1$ was ranked first, while $\vec{X}_3$ was ranked last.\footnote{We also applied the same procedure with the exception that the causal influence is estimated using GC instead of DI. The resulting ranking was $\vec{r} = [2,1,4,3]$, exemplifying the fact that in contrast to DI, GC did not capture the non-linear interactions.} 
\textcolor{\editColor}{We emphasize that similar results for this scenario were reported in \cite{Malladi2016}, but \cite{Malladi2016} used $N=10^5$ samples, while we use $N=2000$, since we did not condition on other sequences.}

\vspace{-0.15cm}
\section{Ranking Stock Indices} \label{sec:methods}

\subsection{Description of the Analyzed Data}

We examine nine major world stock indices taken from three economic regions: North America, Europe, and East Asia. In North America, we examine the Dow Jones Industrial Average (DJI), the NASDAQ Composite (NDX), and the S\&P 500 (SPX), all from the United States. In Europe, we examine the DAX (DAX) from Germany, the CAC 40 (CAC) from France, and the IBEX 35 (IBEX) from Spain. In Asia we examine the Hang Seng Index (HSI) from Hong Kong, the KOSPI Index (KOSPI) from South Korea, and the Nikkei 225 (NKY) from Japan. Each of these indices tracks a weighted sum of stock prices, usually with the intention of capturing the overall movement of their respective exchanges and, more broadly, of the economy of their respective countries. 

We analyzed two types of data: 
\begin{itemize}
	\item 
		Opening and closing prices sampled each day for all nine indices, for a 24 year period starting at the beginning of 1993 and ending at the end of 2016. 
	\item
		Minute by minute prices of three European indices for the year of 2016. These three European exchanges have closely aligned market hours, making them an ideal set for study at the minute-to-minute scale (see the discussion regarding time offsets in the sequel).
\end{itemize}

\noindent All data was obtained using a Bloomberg Terminal. 

\subsection{Pre-Processing of Stock Indices Data} \label{subsec:fin_preProc}

Direct analysis of the raw indices is problematic since financial time-series are not considered to be stationary \cite{Marschinski2002}, and there is a time offset between the different indices, e.g., the trading in Hong Kong ends before the trading in the US begins. The following pre-processing addresses these issues. 

{\bf {\slshape Stationarity}:} It is commonly assumed that applying a simple transformation on the raw signals leads to an asymptotically stationary time-series \cite{Dimpfl2013,Marschinski2002}. Such transformations are the increment and the return given by:
\vspace{-0.1cm}
\begin{align}
		x_{\text{inc},n} = x_{n+1}-x_n, \qquad x_{\text{ret},n} = \frac{x_{n+1}-x_n}{x_n}.
\end{align}

\vspace{-0.1cm}
\noindent Here $x_n$ refers to the closing price of an index at the end of a day or at the end of a minute. In accordance with \cite{Marschinski2002}, we observed that both metrics produced similar results. Hence, in the following we only present the results obtained using the increment transformation $x_{\text{inc},n}$ where, prior to estimating the DI, we removed the mean of the series of increments and normalized its variance to unity. 

{\bf {\slshape Time offset}:} 
%
%
When examining indices from different regions of the world, we offset the time-series {\em to match the order in which exchanges open and close during a trading day}, \textcolor{\changeColor}{similar to \cite{Kwon2008}}. For example, let \textcolor{\changeColor}{$A_{1}^{N}$ and $U_1^N$} denote the series of increments corresponding to indices from \textcolor{\changeColor}{Asia} and the \textcolor{\changeColor}{US}, respectively. Then, to account for the time offset we estimated
\vspace{-0.3cm}
\textcolor{\changeColor}{
\begin{align*}
I(\vec{A} \to \vec{U}) &= \lim_{N \to \infty} \frac{1}{N} I(A_{2}^{N} \to U_1^{N-1}), \\
I(\vec{U} \to \vec{A}) &= \lim_{N \to \infty} \frac{1}{N} I(U_{1}^{N} \to A_1^N).
%
\end{align*} 
}

\vspace{-0.2cm}
\noindent \textcolor{\changeColor}{\noindent Several studies ignore this time offset between examined markets, considering all closing prices as if they were simultaneously sampled.}
However, we argue that adding this offset correctly accounts for the effect of the earlier-trading exchanges on the later-trading ones within a given day, with respect to Coordinated Universal Time. Because we estimate all DI in a pair-wise fashion, we never need to offset by more than one sample. We consider markets on the same continent (Asia, Europe, or North America) to be simultaneously sampled due to the complete or high overlap of trading periods. Furthermore, we ignore overlap \textcolor{\changeColor}{of US and European} trading hours when ordering indices. Thus, we use the time ordering Asia $\rightarrow$ Europe $\rightarrow$ North America.

In addition to the different trading times, the trading days in the different markets do not fully overlap (for example, due to national holidays). To account for this aspect, for each pair of indices, we removed any dates not common to both.\footnote{Removing dates for pairs rather than globally allows \textcolor{\changeColor}{retention of} a high percentage of the data points.} We found this method preferable to interpolation \cite{Dimpfl2013} \textcolor{\changeColor}{or repetition \cite{Sandoval2015},} as these methods impose additional structure on the data.

\section{Results and Discussion} \label{sec:Results}

We start by examining daily data from January 1993 through December 2016, with a focus on the twelve year period from January 2005 through December 2016. Similarly to \cite{Dimpfl2014}, we found the DI flow between indices to be higher during and following the financial crisis of 2007-2008. Specifically, the mean DI between indices during this time period was found to be 83\% higher than the mean DI from 1993 through 2004, making for easier analysis. The analyzed minute-by-minte data is from January 2016 through December 2016.

\subsection{Estimated Markov Orders} \label{sec:ResultsMarkovOrders}

We estimated the Markov order $M_j$ as discussed in Subsection \ref{subsec:estDI}, where the set of candidate orders was chosen to be $\mathcal{M}_j = \{1,2, ... 5\}$.
Table \ref{tbl:EstMarkovOrders} details the frequency of the estimated Markov orders for daily-sampled data for the time period 1993 - 2016 and the subset of interest, 2005-2016. 
\textcolor{\changeColor}{The table indicates that the sequence of increments has a short memory at the daily scale, with a median Markov order of $1$ and means of 1.33 and 1.39 respectively}
%
The short memory assumption was used in many previous studies, e.g., \cite{Billio2012,Kwon2008,Sandoval2015,Marschinski2002} all assume a memory order of $1$ in their calculations. Our Markov order estimation corroborates this assumption.

\begin{table}[h]
\captionsetup{font=footnotesize}
\caption{Number of times a Markov order was estimated.}
\label{tbl:EstMarkovOrders}
\vspace{-0.3cm}
\begin{center}
\begin{tabular}{|c|c|c|c|c|c|}
\hline																																						
Period & $M_j \mspace{-3mu} = \mspace{-3mu} 1$ & $M_j \mspace{-3mu} = \mspace{-3mu} 2$ & $M_j \mspace{-3mu} = \mspace{-3mu} 3$ & $M_j \mspace{-3mu} = \mspace{-3mu} 4$ & $M_j \mspace{-3mu} = \mspace{-3mu} 5$  \\
\hline
\hline
1993 -- 2016 & 61 & 10 & 1 & 0 & 0 \\
\hline
2005 -- 2016 & 57 & 13 & 1 & 1 & 0 \\
\hline
\end{tabular}
\end{center}
\vspace{-0.3cm}
\end{table}

Next, we discuss the estimated DI values.

\subsection{Estimated Directed Information Values}

\begin{figure}[t]
\captionsetup{font=footnotesize}
\begin{center}
\includegraphics[width=0.70\columnwidth]{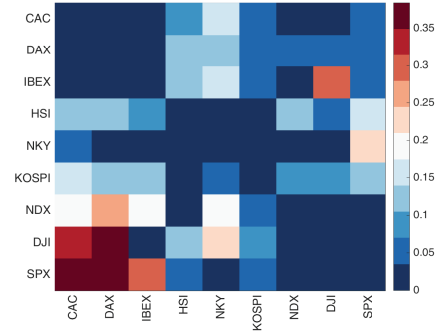}
\end{center}
  \vspace{-0.2cm} 
	\caption{Heat map of estimated DI values for the period 2005-2016.}
\label{fig:heatmap}
\vspace{-0.4cm}
\end{figure}

We estimated the DIs between each pair of index time-series. The heat map in Fig. \ref{fig:heatmap} depicts these estimations, where the $(i,j)$ entry corresponds to the DI from index $i$ to index $j$. 
It can be observed that the DI estimated between indices from {\em the same region}, depicted in the 3 by 3 blocks along the diagonal, is very small. This suggests that information flow {\em inside a region} might occur at a much faster pace, and daily sampling is insufficient to capture the intra-region information flow.
Motivated by this observation, Fig. \ref{fig:DISuperNodes} depicts a directed graph where each node correspond to a {\em region}, and the weight of the edge from node $i$ to node $j$ equals the {\em sum of the estimated DIs} from region $i$ to region $j$. 
As expected, Fig. \ref{fig:DISuperNodes} shows a high amount of causal influence from the US indices to other global regions, in particular to Europe. 
A similar phenomena was discussed in \cite{Kwon2008}, though in \cite{Kwon2008} the causal influence from the US to Asia is higher than the one from the US to Europe. Use of a different set of indices may explain this discrepancy, as the analysis only captures DI within this set. 

\tikzset{
    vertex/.style={circle,draw,minimum size=5em},
    edge/.style={->,> = latex'}
}
\tikzset{
    vertex/.style={circle,draw,minimum size=6em},
    edge/.style={->,> = latex'}
}
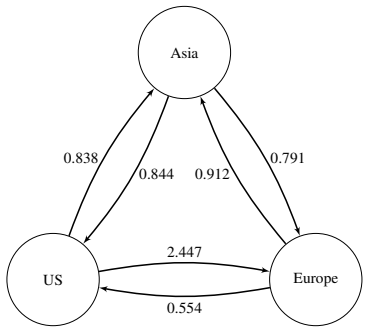
\begin{figure}[t]
\captionsetup{font=footnotesize}
\begin{center}
\resizebox {.55\columnwidth} {!}
{
\begin{tikzpicture}

  \node[vertex] (1) at (3,0) {Europe};
  \node[vertex] (3) at (-3,0) {US};
  \node[vertex] (2) at (0,5.19) {Asia};

  
  \draw[edge, line width=1pt] (1) edge [bend left = 10] node[midway, below] {0.554} (3);
  \draw[edge, line width=1pt] (1) edge [bend left = 10] node[midway, left] {0.912} (2);
  \draw[edge, line width=1pt] (2) edge [bend left = 10] node[midway, right] {0.791} (1);
  \draw[edge, line width=1pt] (2) edge [bend left = 10] node[midway, right] {0.844} (3);
  \draw[edge, line width=1pt] (3) edge [bend left = 10] node[midway, above] {2.447} (1);
  \draw[edge, line width=1pt] (3) edge [bend left = 10] node[midway, left] {0.838} (2);

\end{tikzpicture} }
			\vspace{0.1cm}
			\caption{Estimated DI graph between the regions: US, Asia, and Europe \textcolor{\changeColor}{with daily sampling}.}
      \label{fig:DISuperNodes}
\end{center}
\vspace{-0.65cm}
\end{figure}

Closely examining Fig. \ref{fig:heatmap}, one can find {\em a significant amount of DI from IBEX to DJI and a small amount of DI from DJI to IBEX}. As the economy of Spain is relatively small compared to the US, this result is somewhat counter-intuitive. Thus, we investigated this phenomenon further, as we next discuss in more detail.

\subsection{Spain as an Indicator of the European Union} \label{sec:IBEXDJI}
\begin{figure}[t]
\captionsetup{font=footnotesize}
\begin{center}
\includegraphics[width=0.65\linewidth]{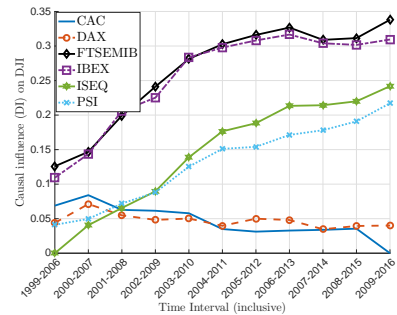}
\end{center}
\vspace{-0.25cm} 
   \caption{DI from CAC, DEX, FTSEMIB, IBEX, ISEQ, and PSI to the DJI; estimated over an eight-year window.}
\label{fig:IBEXDJI}
\vspace{-0.55cm} 
\end{figure}

\textcolor{\changeColor}{
While Fig. \ref{fig:heatmap} shows a high DI from the IBEX to the DJI, this relationship is reversed for the other two examined European indices, which have much larger economies than Spain. Thus, this result conflicts with the intuition that larger economies generally influence smaller economies. 
Upon discovery of this relationship, we first investigated the change of $I(\text{IBEX} \to \text{DJI})$ over time. The line {IBEX} in Fig. \ref{fig:IBEXDJI} indicates $I(\text{IBEX} \to \text{DJI})$ over time, where each point is estimated with an eight-year window to retain a sufficient sample size. We observe a large increase in influence between 2007 and 2012.
}

\textcolor{\changeColor}{
We conjecture that the increase in causal influence between the IBEX and the DJI is related to the 2007--2008 financial crisis and the subsequent recession. 
We believe that this result is due to Spain's economic situation during the crisis.
After the 2007--2008 financial crisis, several European countries, specifically Portugal, Ireland, Italy, Greece, and Spain,\footnote{These countries are sometimes referred to as PIIGS \cite[p. 229]{Quiggin2012}.} were unable to independently weather the recession without the European Union's support. 
Spain's credit rating was also cut several times during this period. Whereas Greece was a small economy whose default did not represent a significant threat to the euro, Spain had a much bigger economy and, as a result, was a much bigger threat. If Spain failed, it would cause significant economic distress in the European Union. 
Therefore, we hypothesize that {\em Spain's status as an indicator} of the future stability of the euro and European Union might have caused the Spanish IBEX index to exert undue influence on the DJI.
}

\textcolor{\changeColor}{
To test this hypothesis, we examined the influence on the DJI of a stock index from Italy (FTSEMIB), a {\em similarly sized economy} which was also at risk during the financial crisis. If the above analysis is correct, one should expect $I(\text{FTSEMIB} \to \text{DJI})$ to behave similarly to $I(\text{IBEX} \to \text{DJI})$. In addition, we also examined the DI from the Portuguese (PSI) and Irish (ISEQ) indices. Fig. \ref{fig:IBEXDJI} depicts the corresponding estimated DIs over an eight-year sliding window. We see that the influence on the DJI of all indices from at-risk economies significantly increases during this time period, and the DI measures from the Spanish and Italian are highly correlated. 
We conjecture that since Portugal and Ireland have economies smaller than Spain and Italy, they are weaker indicators of the financial status of the European Union. 
In contrast to the influence of Spain, Italy, Portugal and Ireland, we see that the influence of the French and German indices on the DJI does not significantly change, and is relatively low. We believe that this is because there was no true fear that either France or Germany was in danger of an economic meltdown that would significantly impact global markets.
}

\textcolor{\changeColor}{
Note that while we see a significant influence from the IBEX to the DJI, we don't see the same for the other examined US indices. We conjecture that this relates to the composition of the indices, but this conjecture requires exploration which we leave to future work.
}

\vspace{-0.15cm}
\subsection{Ranking Results} \label{sec:resultsCI}

To rank the different indices in terms of their mutual causal influence we applied the approach \textcolor{\changeColor}{discussed in Section} \ref{sec:proposedMethod}. The resulting net-flow (sum of incoming flow subtracted from the sum of outgoing flow) is detailed in Table \ref{tbl:pairwiseDI_NF}.


\begin{table}[h]
\captionsetup{font=footnotesize}
\caption{The Net-flow metric for the nine considered indices.}
\label{tbl:pairwiseDI_NF}
\vspace{-0.3cm}
\begin{center}
\begin{tabular}{|c|c||c|c||c|c|}
\hline																																						
{\bf Index} & {\bf Netflow} & {\bf Index} & {\bf Netflow} & {\bf Index} & {\bf Netflow}  \\
\hline
\hline
Dow Jones & 0.699 & KOSPI & 0.317 &  Nikkei & -0.614\\
\hline
Nasdaq & 0.658 & Hang Seng & 0.182 & DAX & -0.827 \\
\hline																																																																				
S\&P 500 & 0.530 & IBEX & -0.068 &  CAC & -0.877 \\
\hline
\end{tabular}
\end{center}
\vspace{-0.3cm}
\end{table}


It can be observed that Table \ref{tbl:pairwiseDI_NF} ranks the US indices as the most influential; this is consistent with the intuition that the indices representing countries with larger economies usually exert more influence than the indices representing countries with smaller economies, and similar indices should be clustered together. We see that the US and Chinese indices are all net sources of information according to the net-flow algorithm. However, France and Germany sit at the bottom of the ranking, whereas Spain and Korea, with smaller economies, sit in the middle of the ranking.\footnote{Note that the intra-European influence inferred based on the minute-by-minute sampling implies a different ranking among the European indices.} 
We must keep in mind that this selection of indices represents a small subset of all global indices, and analysis in a closed environment overlooks many significant global interactions.

\begin{figure}[t]
\captionsetup{font=footnotesize}
\begin{center}
\includegraphics[width=0.75\linewidth]{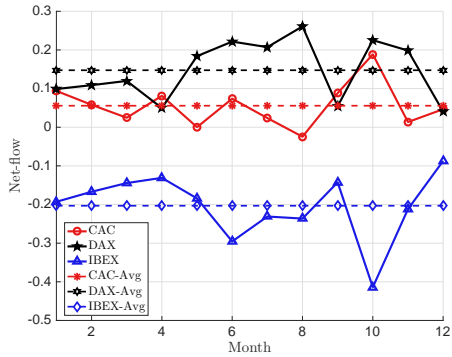}
\end{center}
\vspace{-0.25cm} 
   \caption{Monthly net-flow calculated for the European indices, for minute-by-minute sampling.}
\label{fig:netflowEurope}
\vspace{-0.55cm} 
\end{figure}

We also calculated the net-flow metric for the DI graph produced from the minute-by-minute sampled data for the three European indices. 
We repeated the calculation for each month-long block of data (corresponding to the calendar months). Fig. \ref{fig:netflowEurope} depicts the resulting net-flow values. Within this region, the ranking of countries also corresponds to the size of their economy, with the German index, on average, having the highest net-flow, followed by the French index, then the Spanish index. Note that the German and French indices have a net outflow of information, whereas the Spanish index has a net inflow of information. 

Overall, the computed influence largely corresponds to our intuition as to who affects whom in the global economy. Without an underlying truth as to the behavior of global markets, precisely quantifying the accuracy of these results is difficult.

\vspace{-0.25cm}
\section{Conclusion} \label{sec:conclusion}

\vspace{-0.1cm}
We presented a non-parametric method for ranking a set of stock indices according to their causal influence. 
Daily as well as minute-by-minute data from nine indices was analyzed.
In general, our results are consistent with the intuition that the indices representing countries with larger economies usually exert more influence than indices representing countries with smaller economies.
Our analysis showed that an exception is the influence from the IBEX to the DJI, which is conjectured to follow from the fact that Spain serves as an indicator for the European Union financial status. 
Under the assumption that indices reflect the underlying economy of a country, the proposed ranking indicates which countries exert the most economic influence in the examined subset of the global economy.


\bibliographystyle{IEEEtran}
\vspace{-0.15cm}
\bibliography{IEEEabrv,references}

\end{document}